# First Thin-Film Lithium Tantalate Polarization Controller Enabling Reset-Free Mrad·s⁻¹ Tracking for Optical Interconnects

Zichao Gao[1,†], Siyu Lu[1,†], Mingming Zhang[1,†], Gengqi Yao[1], Chicheng Zhang[1], Miao Deng[1], Siyu Chen[1], Yiqi Dai[1], Shiqi Yue[1], Chijun Li[1], Yuqi Li[1], Ziwen Zhou[1], Zheli Liu[1], Xinyang Yu[1], Xitao Ji[1], Cheng Zeng[1,*], Siqi Yan[1,2,*], Jinsong Xia[1,*] and Ming Tang[1,2,*]

[1] School of Optical and Electronic Information and Wuhan National Laboratory for Optoelectronics, Huazhong University of Science and Technology, Wuhan 430074, China

[2] Hubei Optical Fundamental Research Centre, Wuhan 430074, China

*Corresponding author: Cheng Zeng. Email: zengchengwuli@hust.edu.cn, Siqi Yan. Email: siqya@hust.edu.cn, Jinsong Xia. Email: jsxia@hust.edu.cn and Ming Tang. Email: tangming@mail.hust.edu.cn

† These authors contributed equally to this work.

## Abstract

The rapid escalation of computing power driven by large-scale artificial intelligence is placing unprecedented demands on the bandwidth, latency, and energy efficiency of data-center interconnects (DCIs). Self-homodyne coherent (SHC) transmission is a promising architecture because it preserves the spectral efficiency of coherent detection while greatly simplifying digital signal processing, but its practical deployment is critically limited by random and often ultrafast state-of-polarization (SOP) fluctuations that induce carrier fading and destabilize coherent reception. Here we report the first integrated polarization controller based on thin-film lithium tantalate (TFLT), enabling reset-free polarization tracking at Mrad·s⁻¹ speeds. The four-stage electro-optic device exhibits polarization-dependent loss (PDL) below 0.3 dB, a half-wave voltage below 2.5 V, high modulation bandwidth, and negligible DC drift. To accommodate the finite tuning range of integrated phase shifters, we develop a finite-boundary gradient-descent (FBGD) control algorithm that ensures reset-free SOP evolution with no phase jump. The implemented adaptive polarization controller (APC) is validated through both standalone polarization-tracking measurements and a dual-polarization 16-QAM SHC 400-Gbps transmission system. Transient polarization disturbances can be tracked at speeds up to 2 Mrad·s⁻¹, while stable reset-free operation under continuous polarization disturbances is maintained up to 1 Mrad·s⁻¹. This reset-free performance represents more than doubling the state of the art, while the pre-FEC bit-error rates remain below the HD-FEC threshold under realistic DCI conditions and lightning-scale polarization disturbances. These results establish TFLT as a new platform for ultrafast, low-power, reset-free, and drift-free polarization control in coherent optical interconnects and beyond.

## Introduction

The rapid escalation of computing power driven by large-scale artificial intelligence models is fundamentally redefining the requirements of data-center interconnects (DCIs)[1–3]. As modern AI training and inference increasingly rely on distributed clusters of GPUs and accelerators, system-level performance is no longer limited by the speed of individual processors, but by the capacity, latency, and energy efficiency of the optical links that connect them[4–6]. While coherent optical transmission offers the spectral efficiency needed to sustain this growth, its reliance on complex digital signal processing and high-power local oscillators conflicts with the tight power and cost budgets of large-scale computing fabrics[6]. Self-homodyne coherent (SHC) architectures have therefore emerged as an attractive alternative, preserving the capacity advantage of coherent detection while eliminating carrier recovery and dramatically simplifying signal processing[6–8]. This architectural simplification, however, shifts a critical burden from the digital to the optical domain: SHC systems become intrinsically sensitive to polarization dynamics[9].

In an SHC receiver, coherent demodulation requires stable alignment between the SOP of the signal and that of the transmitted carrier. Any mismatch directly reduces the effective local-oscillator power, leading to carrier fading and, in the worst case, complete loss of coherent detection[10]. In deployed fiber links, SOP fluctuations are continuously induced by birefringence, mechanical vibration, and temperature drift. While typical polarization rotation rates in data-center environments are relatively slow, extreme transient events such as lightning strikes, power switching, or strong electromagnetic interference coupling into fiber spans can drive polarization dynamics into the hundreds of krad·s⁻¹ or even the Mrad·s⁻¹ regime[11,12]. Such ultrafast excursions lie far beyond the tracking capability of existing integrated polarization controllers and represent a worst-case but realistic threat to the stability of SHC links, where even brief loss of polarization alignment can interrupt carrier recovery and disrupt data transmission[6], an event that in large-scale AI computing centers can stall distributed training jobs across thousands of GPUs, leading to cascading synchronization failures[13] and potentially incurring losses of hundreds of thousands of dollars from a single interruption.

Moreover, this sensitivity to polarization is not unique to SHC systems. Dual-polarization intensity-modulation and direct-detection (DP-IMDD) architectures, which are rapidly gaining traction as a low-cost solution for short-reach and data-center links [14,15], likewise depend on stable and well-defined polarization states to preserve channel orthogonality and suppress inter-polarization crosstalk. In these systems, uncontrolled state-of-polarization (SOP) drift directly degrades signal integrity and limits the scalability of polarization-multiplexed IMDD, making fast and robust optical-domain polarization control a universal requirement across both coherent and direct-detection interconnects.

Despite this urgent need, current integrated polarization-control technologies remain constrained by fundamental trade-offs of the materials. Silicon photonic platforms offer excellent scalability and manufacturing maturity[16], but rely on thermo-optic or carrier-based phase shifters whose intrinsically slow response and parasitic intensity modulation fundamentally limit both the bandwidth and fidelity of polarization tracking[17]. Thin-film lithium niobate enables fast electro-optic phase modulation, yet practical operation typically requires thermal bias control to compensate DC drift[18,19], introducing a slow control loop that conflicts with continuous and rapidly varying SOPs. As a result, no existing platform can simultaneously provide high-speed tracking, low power consumption, and long-term drift-free stability, all of which are indispensable for practical SHC-based DCI systems.

Beyond material limitations, polarization control itself is subject to a more fundamental constraint arising from the finite phase space of the integrated phase shifter. Because integrated phase shifters have a finite tuning range, continuous SOP evolution inevitably drives the control variables to their boundaries, forcing abrupt phase resets, for example from $2\pi$ to $0$[20]. These resets cannot occur instantaneously and must sweep through intermediate phase states, during which polarization tracking is lost and coherent reception collapses. Previous strategies, such as constraining the SOP to circular trajectories on the Poincaré sphere[21] or employing optical state switching[22], can reduce the frequency of these events but do not eliminate them. Truly reset-free polarization tracking under continuous and high-speed SOP rotation therefore remains an open challenge.

Here we overcome these intertwined material and algorithmic barriers by introducing the first integrated polarization controller based on thin-film lithium tantalate (TFLT). TFLT combines low optical loss with a strong electro-optic response and, critically, exhibits intrinsically low DC drift under low-frequency driving, enabling stable bias-free operation. Building on these properties, we develop a finite-boundary gradient-descent control algorithm that enables continuous, reset-free polarization tracking despite the finite tuning range of integrated phase shifters. We validate the TFLT-based adaptive polarization controller (APC) through both standalone SOP-tracking measurements and a 400-Gbps 16-QAM SHC transmission system. Our device achieves polarization-tracking speeds up to 2 Mrad·s$^{-1}$ for transient polarization disturbances and stable reset-free operation under continuous polarization disturbances is maintained up to 1 Mrad·s$^{-1}$, representing the fastest integrated polarization tracking reported to date and covering virtually all polarization dynamics encountered in practical fiber links, from slow environmental drifts to extreme, lightning-scale perturbations, while preserving stable coherent reception throughout. By simultaneously delivering ultrafast response, low polarization-dependent loss (PDL), low driving voltage, and drift-free operation, this work establishes a new regime of polarization control, within which multiple key performance metrics extend beyond the current state of the art, unlocking the practical scalability of both SHC and DP-IMDD optical interconnects for next-generation AI-driven data-center networks.

## Results

**Device Design**

We demonstrate a TFLT APC on an x-cut lithium tantalate-on-insulator (LTOI) substrate with a 400-nm-thick lithium tantalate layer. The device is based on three fundamental building blocks: edge couplers (EC), a polarization splitter-rotator (PSR), and 2×2 Mach-Zehnder interferometers (MZI), composed of four EO phase shifters and four 3-dB multi-mode interferometer (MMI) couplers, as shown in Figure 1a. The phase differences between the two orthogonal electric field components $E_x$ and $E_y$, which constitute an arbitrary SOP, are reconfigured by the four EO phase shifters.

The principle of polarization locking is illustrated here. The received signal $E_{out1}$ and $E_{out2}$ at two output ports can be written as follows:

$$\begin{bmatrix} E_{out1} \\ E_{out2} \end{bmatrix} = \frac{1}{2}\begin{bmatrix} 1 & j \\ j & 1 \end{bmatrix}\begin{bmatrix} e^{j\varphi_2} & 0 \\ 0 & 1 \end{bmatrix}\begin{bmatrix} 1 & j \\ j & 1 \end{bmatrix}\begin{bmatrix} e^{j\varphi_1} & 0 \\ 0 & 1 \end{bmatrix}\begin{bmatrix} E_x \\ E_y \end{bmatrix}$$
$$= je^{j\frac{\varphi_2}{2}}\begin{bmatrix} \sin(\frac{\varphi_2}{2}) & \cos(\frac{\varphi_2}{2}) \\ \cos(\frac{\varphi_2}{2}) & -\sin(\frac{\varphi_2}{2}) \end{bmatrix}\begin{bmatrix} e^{j\varphi_1}E_x \\ E_y \end{bmatrix} \qquad (1)$$

where $E_x$ and $E_y$ correspond to a pair of orthogonal polarization states, $\varphi_1$ and $\varphi_2$ are the phase delays of two arms (two-stage EO phase shifters considered here), and $\begin{bmatrix} 1 & j \\ j & 1 \end{bmatrix}$ is the transmission matrix of a 3-dB 2×2 MMI. Using Eq.(1), a specific set of parameters [$\varphi_1$, $\varphi_2$] can be derived to maximize the optical power at one output port while minimizing it at the opposite port as a monitor. While a two-EO phase shifter architecture is sufficient for polarization tracking, its operation requires a reset process once the driving voltages approach their limits, causing transient signal interruptions. To avoid this limitation, a device incorporating four-stage EO phase shifters is adopted.

Figure 1b illustrates the cross-sectional schematic of the EO phase shifter. The ridge waveguide, oriented along the y-axis of the LT crystal, features a ridge height of $H_r$ = 240 nm and a width of $W_m$ = 4 µm. Additionally, gold (Au) electrodes with a height of $H_e$ = 1.2 µm are directly deposited onto the LT slab. The EO phase shifters are designed in a single-drive push-pull configuration to maximize modulation efficiency, featuring an electrode gap of $D_m$ = 6 µm. Figure 1c depicts the corresponding simulation of the optical-electrostatic field interaction within the modulation section.

The detailed schematic of the polarization splitter-rotator (PSR) is illustrated in Figure 1d. The PSR comprises an adiabatic taper and an asymmetric directional coupler (ADC). In the adiabatic taper section ($L_1$ = 380 µm), the waveguide width is designed to vary from $W_0$ = 0.9 µm to $W_1$ = 1.7 µm. This section rotates and converts the input TM0 mode into the TE1 mode, while the TE0 mode propagates directly to the through port without conversion. Subsequently, the TE0 and TE1 modes are separated into different output ports via the ADC. In the ADC region, the width of the through waveguide decreases linearly from $W_1$ to $W_2$ = 1.5 µm, while the width of the cross waveguide increases linearly from $W_3$ = 0.3 µm to $W_4$ = 0.8 µm. The gap between the two waveguides remains constant at $W_g$ = 0.6 µm. The lengths of the three ADC regions ($L_2$, $L_c$, $L_3$) are set to 80, 180, and 140 µm, respectively. Figures 1f and 1g present the calculated transmission spectra for TM0, TE0 and TE1 modes at the output ports over a bandwidth of 1500–1620 nm. The TE0 mode maintains near-unity transmission with an extinction ratio (ER) exceeding 40 dB.

Conversely, for the TM0 mode input, the insertion loss at the cross port is less than 0.05 dB. At the through port, crosstalk is suppressed below -30 dB for the TM0 mode and below -24 dB for the TE1 mode. Consequently, the proposed PSR demonstrates negligible PDL, low insertion loss and low channel crosstalk over a broad bandwidth.

Figure 1e depicts the edge coupler, which employs a bilayer taper structure to ensure low loss and minimal PDL between the optical fiber and the LTOI chip. The structure, comprised of a silicon oxynitride (SiON) waveguide, a bottom waveguide layer, and a top waveguide layer, is segmented into three adiabatic transition regions to facilitate efficient mode evolution. In the first adiabatic stage, the input mode is confined within the bottom waveguide, where the width expands from $r_1 = 0.1$ μm to $r_2 = 1.8$ μm to enlarge the mode field. The critical mode conversion occurs in the second section. Here, a taper with a narrow tip width of $t_1 = 0.1$ μm is introduced on the bottom waveguide layer. As light propagates through this region, the optical mode is adiabatically transferred from the bottom layer to the upper layer via simultaneous width variation ($r_2$ to $r_3 = 2.4$ μm and $t_1$ to $t_2 = 1$ μm). Finally, the optical power is fully coupled into the top TFLT waveguide in the last section. Figure 1h shows the simulated transmission performance of the edge coupler over a 120 nm bandwidth, indicating insertion loss less than 0.35 dB for both TE and TM modes, with PDL maintained below 0.3 dB.

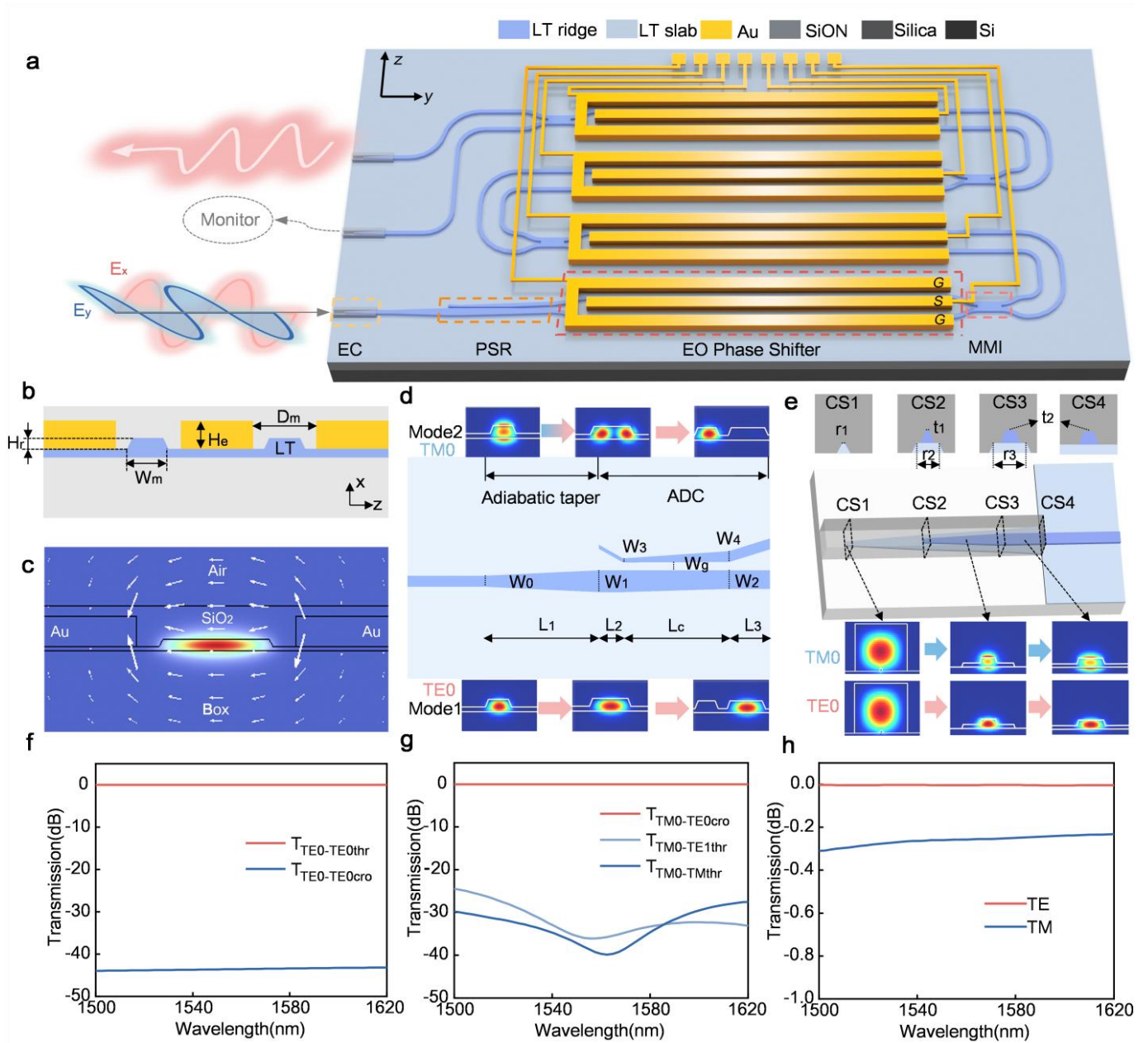

**Figure 1. Device architecture and key building blocks of the TFLT-based APC. a** Schematic configuration of the TFLT APC. **b** The cross-section of the EO phase shifter. **c** Simulated the optical mode profile and electrostatic field vector distribution in the modulation section. **d** The schematic diagram of the polarization splitter and rotator and **e** the edge coupler. **f, g, h** Simulated transmission spectrum of the PSR for (**f**) TE0 (**g**) TM0 inputs, and (**h**) the edge coupler.

## Fabrication and Characterization

The proposed device was fabricated on a 400-nm-thick x-cut TFLT wafer with a 4.7-μm-thick buried oxide layer (NanoLN). The APC waveguide patterns were defined using electron beam lithography (EBL) and inductively coupled plasma (ICP) dry etching by $Ar^+$ gas, in which way the etch depth was 240 nm and the etched waveguide had a sidewall angle of 70°. The next processes of photolithography, evaporation, and lift-off were used to form patterns of gold electrodes. The edge couplers were fabricated using similar steps, which we have demonstrated for LNOI platforms[23]. We employ a fiber array (FA) coupling to facilitate the test, and electrode encapsulation was achieved by wire bonding to a printed circuit board (PCB). The microscopy picture of the whole device and detailed scanning electron micoscopy (SEM) pictures are shown in Figure 2a.

To establish the intrinsic speed of the TFLT platform, a traveling-wave electro-optic modulator with the same 240 nm etch depth was fabricated and characterized. The measured S-parameters and extracted electro-optic response (Figure 2b) show only a 2.03 dB drop in S21 at 67 GHz, corresponding to a projected 3-dB electro-optic bandwidth of 86 GHz. This confirms that the TFLT waveguide platform supports sub-nanosecond phase modulation and is fundamentally capable of operating in the ultrafast regime required for Mrad·$s^{-1}$ polarization control. For the APC, the phase shifters were deliberately implemented with 1 cm lumped electrodes rather than traveling-wave structures, reflecting a system-level trade-off: polarization dynamics in fiber links are dominated by low-frequency components, and lumped electrodes provide improved bias stability, lower RF complexity, and simpler packaging without limiting the relevant tracking bandwidth. After wire bonding and PCB integration, the packaged 3 dB electro-optic bandwidth is approximately 1 GHz, still orders of magnitude faster than practical polarization fluctuations.

Figures 2c and 2d present the normalized transmission spectra of the PSR and the edge coupler, revealing an ultra-low PDL below 0.3 dB across the entire C- and L-bands. This remarkably low PDL underscores the polarization-insensitive nature of both components. The PSR further provides an extinction ratio exceeding 20 dB over the full measurement range, ensuring efficient separation and recombination of orthogonal polarization components. The complete four-stage APC, incorporating four electro-optic phase shifters, exhibits an on-chip insertion loss below 4 dB within a compact footprint of 1.4 cm × 1.3 mm.

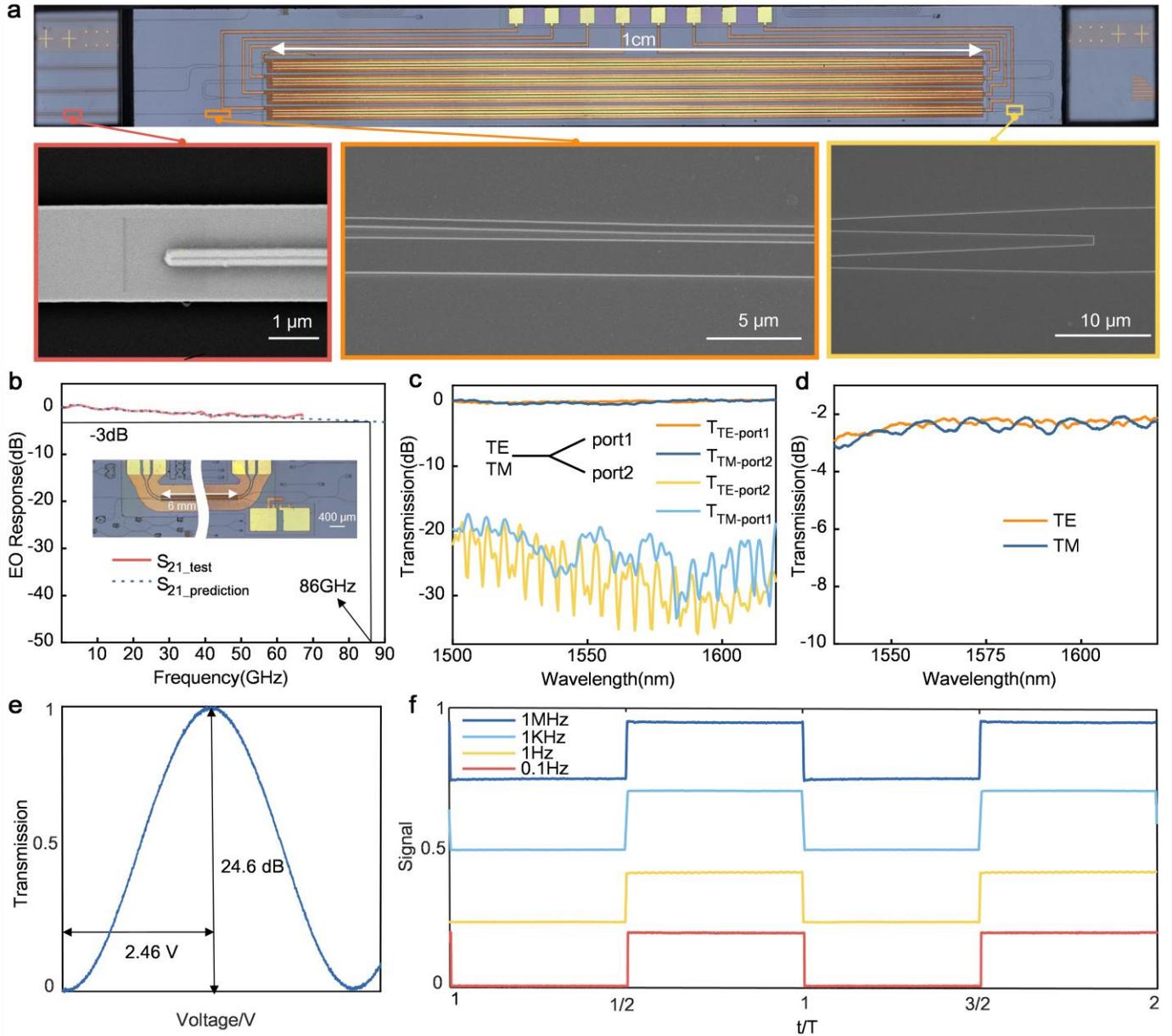

**Figure 2. Fabrication and electro-optic characterization of the TFLT-based APC. a** Microscope image of the fabricated device, and detailed SEM images: edge coupler, PSR, MMI. **b** Measured EO response of the traveling wave electrode modulator. **c,d** Measured transmission spectrum of (**c**) the PSR and (**d**) the edge coupler. **e** Test results of half-wave voltage and extinction ratio for MZI. **f** Response of the device to different square wave frequencies.

Static and dynamic electro-optic characteristics were measured by launching a 1550 nm continuous-wave laser into the chip and driving a single APC stage with a 1 MHz sawtooth waveform. Figure 2e illustrates the measured half-wave voltage of the Mach–Zehnder interferometer (MZI) of 2.46 V, with an ER of approximately 24.6 dB. DC drift is primarily attributed to the migration of charge carriers induced by lattice defects and interfaces within multilayer structures (e.g., $SiO_2$ cladding), which leads to the shielding of the applied electric field[24,25]. Compared to LN, LT exhibits a lower intrinsic defect density and weaker photorefractive effects[26,27]. Furthermore, we optimized the device fabrication process by establishing direct contact between electrodes and LT to eliminate dielectric relaxation paths, alongside improving the $SiO_2$ deposition technique. This synergistic combination of intrinsic material advantages and interface engineering significantly suppresses DC drift. To evaluate dynamic stability, square-wave signals with frequencies ranging from 0.1 Hz to 1 MHz are applied to a single APC stage, as shown in Figure 2f. The output transmission remains stable and fully repeatable over more than seven decades of modulation frequency, indicating negligible DC drift without the need for annealing, which is a critical advantage over most thin-film lithium-niobate devices with slow bias drift typically degrading polarization control[25,28].

**Reset-Free Polarization Tracking via Finite-Boundary Gradient Descent**

A conventional gradient-descent (GD) polarization tracker steers an arbitrary SOP toward a desired output state by iteratively minimizing the detected feedback power[29]. In practical implementations, however, the finite voltage output swing of digital-to-analog converters (DACs) limits the continuous tuning range of the integrated phase shifters in each stage. For randomly varying SOP, a tracking process that relies solely on the GD algorithm will inevitably drive the tuning range to its boundaries. Once a boundary is reached, the corresponding phase must be reset. A commonly adopted approach is to reduce the DAC output of the saturated stage by $2V_\pi$, thereby restoring the available tuning margin for subsequent adjustment. Such a reset strategy, which introduces a $2\pi$ phase jump and forces the APC sweeping through all the intermediate states, inevitably induces discontinuities in the tracking process and can severely degrade the long-term operational performance of the polarization tracker.

To address this issue, we propose a finite-boundary gradient-descent (FBGD) algorithm that enables reset-free polarization tracking under randomly varying SOP. The core idea is to introduce a boundary-aware regularization term into the feedback loss function, which actively and cooperatively adjusts the voltages of the four cascaded phase shifters. By preventing any individual stage from reaching the limits of its tuning range, the proposed approach achieves long-term, stable, and continuous SOP tracking.

From the perspective of finite-resolution digital control, convergence is governed by the local gradient of the loss function. As illustrated in Figure 4a three representative gradient profiles capture the essential behaviors relevant to polarization tracking: a nearly constant gradient, a gradient that weakens near the optimum, and a gradient that increases as the control variable approaches a boundary. For high-speed operation, algorithmic simplicity is essential, because excessive computational complexity directly increases FPGA latency and reduces the achievable control bandwidth. Guided by this constraint, the FBGD algorithm is designed such that the power-minimization term ensures rapid convergence under strong polarization disturbances, while the boundary-regularization term follows the third profile, introducing an increasingly strong restoring force as the phase approaches its limits. This coordinated action enables reset-free tracking without sacrificing convergence speed.

The resulting loss function is written as:

$$L(\boldsymbol{\varphi}) = \alpha P^2(\boldsymbol{\varphi}) + \beta \boldsymbol{\varphi}^2 \tag{2}$$

Where $\boldsymbol{\varphi}$ is the DACs' output voltage vector and $P(\boldsymbol{\varphi})$ is the measured feedback optical power. The weights of the contributions of power minimization and boundary regularization are respectively $\alpha$ and $\beta$. The $i$-th component of the corresponding local gradient when $\boldsymbol{\varphi} = \boldsymbol{\varphi_0}$ is:

$$\left.\frac{\partial L(\boldsymbol{\varphi})}{\partial \varphi_i}\right|_{\boldsymbol{\varphi}=\boldsymbol{\varphi_0}} = 2\alpha P(\boldsymbol{\varphi_0}) \frac{\partial P(\boldsymbol{\varphi_0})}{\partial \varphi_i} + 2\beta \varphi_i \tag{3}$$

And the output of the $i$-th DAC is update iteratively as:

$$\varphi_i^{(n+1)} = \varphi_i^{(n)} - \frac{\partial L(\boldsymbol{\varphi}^{(n)})}{\partial \varphi_i} \tag{4}$$

The effectiveness of the proposed FBGD algorithm is first evaluated by simulation under continuous SOP rotation. As shown in Figure 4b, without regularization the phase distribution accumulates near the physical limits of the phase shifters, indicating a high probability of reset events in practical hardware. When the regularization term is included, the phase distribution is strongly confined away from the boundaries while maintaining comparable tracking accuracy, with the relative intensity error (RIE) of the feedback signal remaining below 0.1[29]. By suppressing boundary accumulation, FBGD enables stable, reset-free polarization tracking under sustained SOP evolution.

In real-time operation, the proposed FBGD algorithm is implemented on an FPGA-based control platform. The feedback optical power is digitized by an analog-to-digital converter (ADC), while the control phase is updated through a DAC driving the phase shifters. The timing sequence of the control loop is illustrated in Figure 4c. With a clock frequency of 200 MHz, the total loop latency, including ADC, DAC, peripheral circuitry, and algorithmic processing, is approximately 100 ns, which provides sufficient bandwidth to support ultrafast polarization tracking under rapid and large-amplitude SOP disturbances.

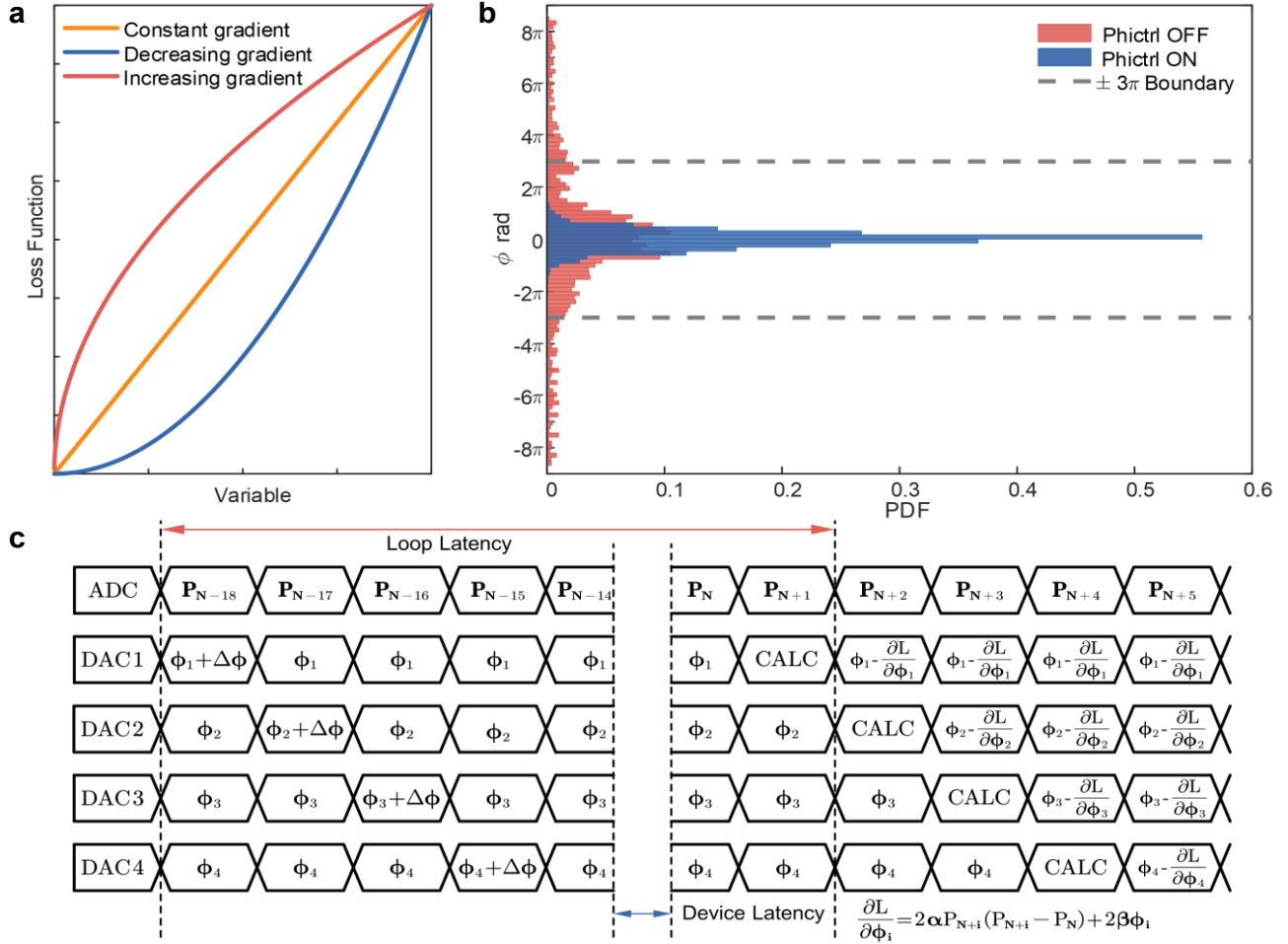

**Figure 4. Finite-boundary gradient-descent algorithm for reset-free polarization tracking. a** Representative gradient profiles of loss functions with a single control variable, illustrating constant, vanishing, and boundary-enhanced restoring gradients. **b** Simulated probability density function (PDF) of the control phase under continuous SOP rotation, comparing conventional gradient descent and the proposed finite-boundary gradient descent. **c** Timing diagram of the FPGA-based FBGD control loop, showing the sequence of sampling, computation, and phase update.

**Reset-free Polarization Tracking of Linear SOP**

Having established both the intrinsic ultrafast electro-optic response of the TFLT platform and the reset-free operation of the FBGD algorithm, we now experimentally probe whether the combined system can sustain polarization tracking in the previously inaccessible Mrad·s$^{-1}$ regime. Figure 5a illustrates the experimental setup for ultrafast reset-free polarization tracking. A continuous-wave laser at 1550 nm is injected into a polarization scrambler that generates arbitrarily rotating SOPs spanning the full Poincaré sphere. The optical signal with distributed SOPs is then injected into the four-stage TFLT APC, whose output is split into a feedback arm and a monitoring arm. The feedback signal is digitized and processed by the FPGA implementing the FBGD algorithm, which updates the four phase shifters through DAC converters in real time, thereby closing a fully optical-electronic polarization control loop.

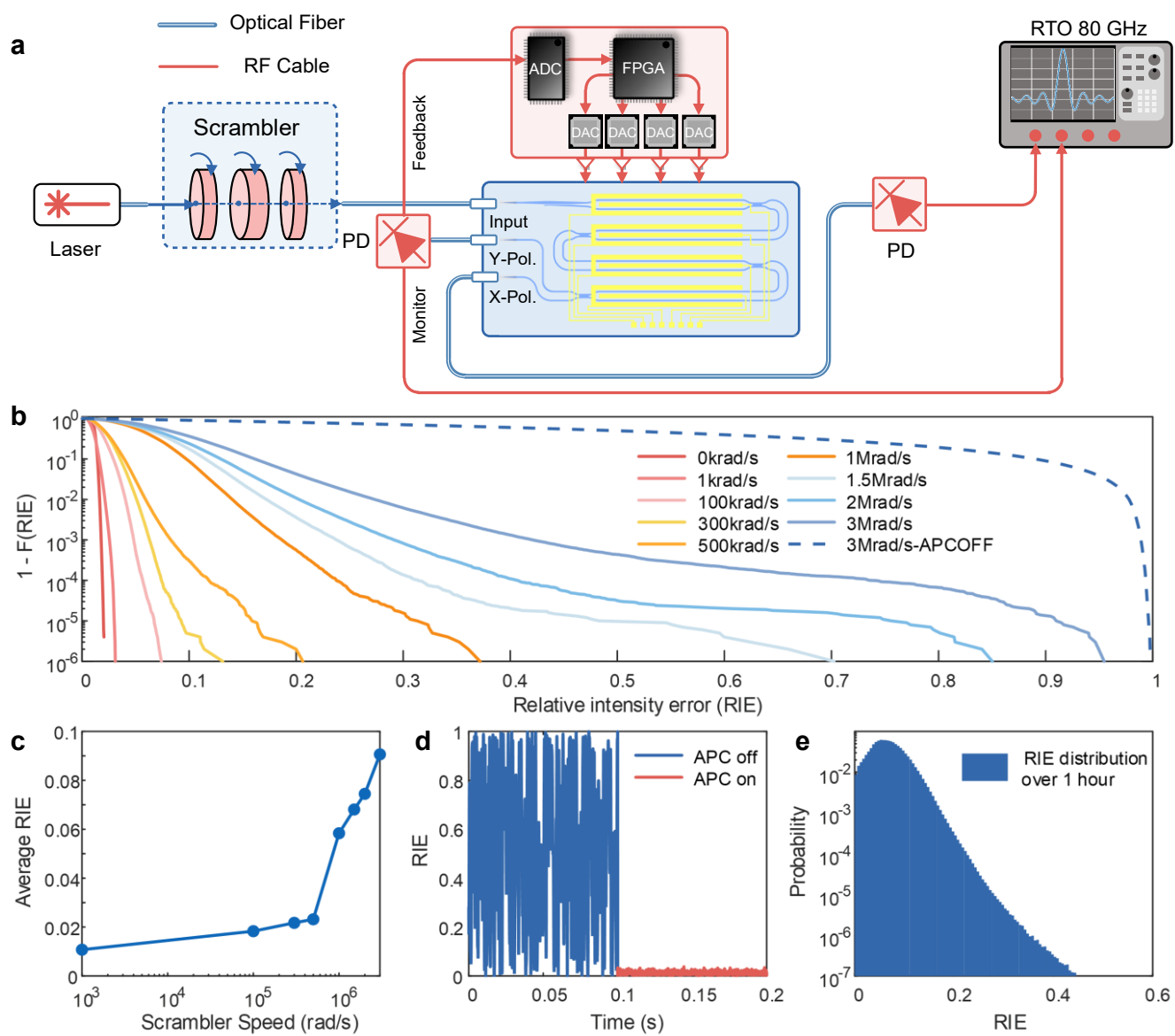

**Figure 5. Reset-free polarization tracking in the ultrafast dynamic regime. a** Experimental setup for characterizing the APC under time-varying input SOPs. **b** Complementary cumulative distribution function (CCDF) of the relative intensity error (RIE), each measured in sampling rate of 1MHz over 10 seconds. **c** The average RIE under different

polarization scrambling speeds, each measured in sampling rate of 1MHz over 10 seconds. **d** Time-domain RIE of the feedback signal with the APC switched off and on, highlighting the effect of active polarization control. **e** The probability distribution of RIE for polarization scrambling speed of 1 Mrad·s$^{-1}$ over 1 hour.

Figure 5b summarizes the polarization tracking performance as a function of scrambling speed using the complementary cumulative distribution function (CCDF) of the RIE, which directly quantifies the residual polarization mismatch. The curves correspond to polarization tracking results obtained over a 10 second interval under SOP scrambling speeds ranging from 0 krad·s$^{-1}$ to 3 Mrad·s$^{-1}$. The leftmost curve represents the reference distribution with the APC enabled in the absence of polarization scrambling, whereas the rightmost curve shows the distribution measured with the APC disabled at a scrambling speed of 3 Mrad·s$^{-1}$, serving as a reference. And Figure 5c shows the average RIE obtained under identical measurement conditions.

For scrambling speeds up to 2 Mrad·s$^{-1}$, the RIE remains well below the lock-loss threshold, with 99.9% of samples below 0.3 and the average RIE is below 0.08, indicates that under transient polarization disturbances, the APC is able to maintain polarization tracking at scrambling speeds of up to 2 Mrad·s$^{-1}$. For scrambling speeds up to 1 Mrad·s$^{-1}$, the RIE remains well below the lock-loss threshold[29,30], with 99.9% of samples below 0.2, and a maximum value below 0.4, demonstrating continuous and reset-free tracking in a dynamic regime that exceeds the capabilities of previously reported integrated controllers[30].

This performance is maintained not only transiently but also over extended operation: Figure 5e shows that the RIE distribution measured over more than one hour at 1 Mrad·s$^{-1}$ is statistically indistinguishable from the short-term results, confirming long-term stability under lightning-scale polarization dynamics.

Under low-speed polarization disturbances relevant to short-reach DCI[31] (<1 krad·s$^{-1}$), as extracted from the average RIE measurements in Figure 5c, the APC maintains an average polarization extinction ratio of 19.6 dB while operating in a reset-free manner, highlighting its ability to span more than six orders of magnitude in polarization-disturbance rate within a single, unified control framework.

**APC Operation in a Self-Homodyne Coherent Link**

To evaluate the performance of the ultrafast polarization control at the system level, we implement a 400-

Gbps 16-QAM SHC transmission system, as illustrated in Figure 6a. The SOP of the local oscillator (LO) light is intentionally disturbed by a polarization scrambler to emulate the rapid and random polarization dynamics encountered in real fiber links. At the receiver end, the modulated signal is detected by an integrated coherent receiver (ICR), and resulting electrical waveforms are captured by an oscilloscope for offline bit-error-rate (BER) analysis. Because SHC performance is directly proportional to the effective LO power delivered to the receiver, any residual polarization mismatch manifests as LO power fluctuations and consequently degrades the BER, thereby providing a sensitive probe of polarization-control fidelity under extreme dynamic conditions.

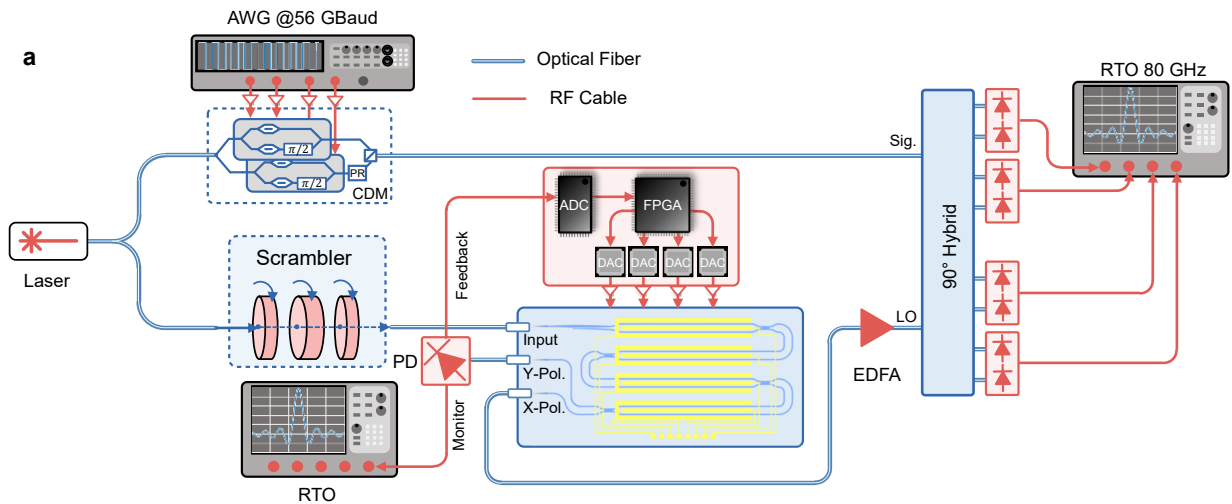

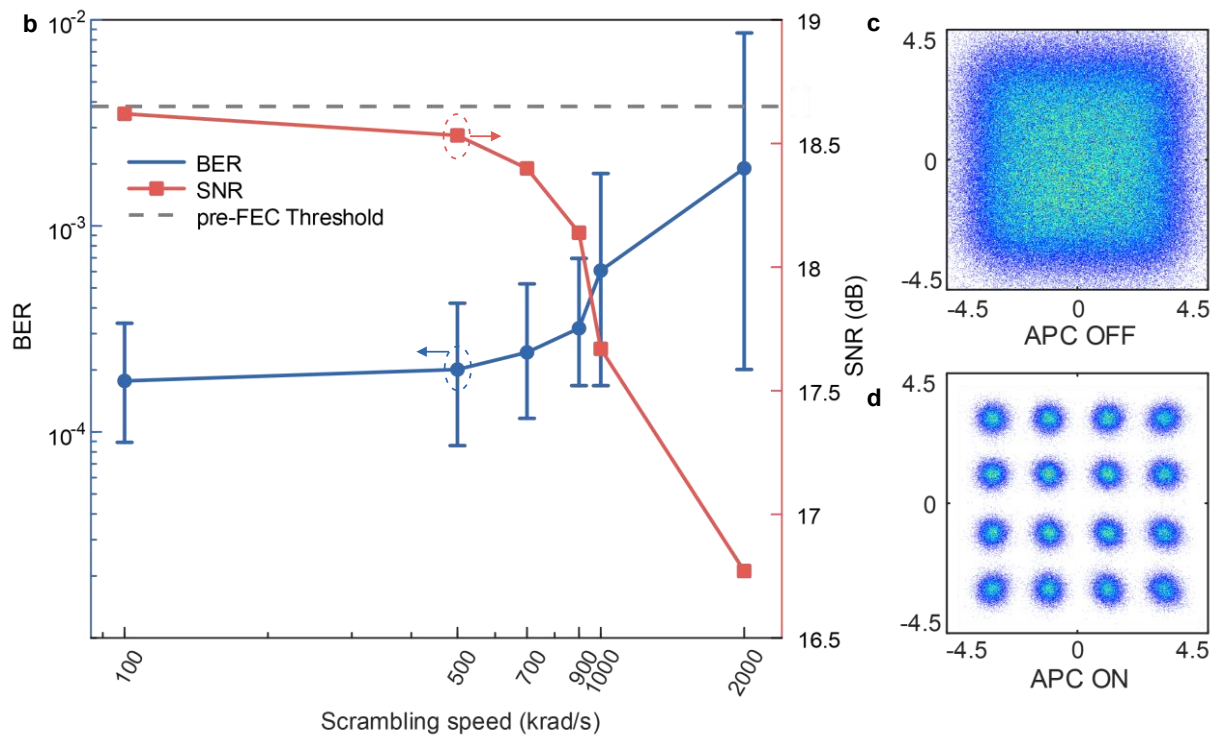

**Figure 6. System-level evaluation of the APC in SHC Link.** **a** Experimental setup to test the APC in SHC communication system. **b** The error bar of the pre-forward error correction BER and the average values of SNR. **c,d** Received constellations with APC disabled or enabled under 1 Mrad·s$^{-1}$ polarization scrambling speed.

Figure 6b summarizes the measured BER performance as a function of scrambling speed. For scrambling speeds ranging from 100 krad·s$^{-1}$ to 1 Mrad·s$^{-1}$, all measured BER values remain below the hard-decision FEC threshold of $3.8 \times 10^{-3}$, demonstrating error-free system operation enabled by reset-free polarization control. Figures 6c and 6d show the received constellations with the APC disabled and enabled under a polarization scrambling speed of 1 Mrad·s$^{-1}$.

When the scrambling speed is increased to 2 Mrad·s$^{-1}$, the BER rises above the FEC threshold, marking the onset of system-level failure. Importantly, this degradation does not originate from a loss of polarization tracking: even at 2 Mrad·s$^{-1}$, 99.9% of the RIE values remain below 0.3 and the average RIE is below 0.08, indicating that the feedback power and SOP alignment are still largely maintained. The observed BER increase instead reflects phase-space exhaustion of the control system: under continuously ultrafast SOP evolution, the finite tuning range of the phase shifters is gradually consumed, preventing the controller from sustaining boundary-free operation. As a result, coherent demodulation is disrupted even though the polarization loop itself remains partially locked. This distinction reveals a fundamental dynamic limit of integrated polarization control and confirms that the performance boundary observed here is not imposed by polarization-tracking fidelity, but by the finite range of practical integrated phase shifters. Crucially, for all realistic fiber perturbations, including extreme polarization dynamics approaching those reported under lightning-induced conditions, the controller operates well below the phase-boundary limit, ensuring stable demodulation and robust link performance. This confirms that the observed breakdown at 2 Mrad·s$^{-1}$ reflects a fundamental extreme-case boundary rather than a practical limitation, and that the proposed TFLT–FBGD scheme remains fully effective across the entire dynamic range relevant to data-center and short-reach optical interconnects.

## Discussion

The central result of this work is not merely a high polarization tracking speed, but the demonstration of continuous, reset-free operation in the megaradian-per-second regime on an integrated photonic platform. While recent lithium-niobate-based and other electro-optic polarization controllers have reported tracking speeds approaching several hundred kiloradians per second, these demonstrations do not address the fundamental instability induced by phase resets under sustained SOP evolution. The experiments presented here show that the key bottleneck for practical polarization control at extreme speeds is not the instantaneous response of the phase shifter, but the ability of the control system to remain within its finite phase space without discontinuous resets.

This distinction becomes clear when the dynamics of the closed-loop control system are analyzed quantitatively. The maximum sustainable polarization rotation rate that can be tracked without phase overflow is governed by the loop latency:

$$\begin{aligned} T_{loop} &= T_{ADC} + T_{DAC} + T_{PD} + T_{Algorithm} \\ &= 10 \times T_{clockcycle} + 11ns + 15ns + 5 \times T_{clockcycle} \end{aligned} \quad (5)$$

which sets the fastest rate at which corrective phase updates can be applied. In the present implementation, $T_{loop} \approx 100$ ns, corresponding to a theoretical tracking bandwidth on the order of 10 MHz. Importantly, this limit arises entirely from electronic and algorithmic delays rather than from the photonic device itself. The TFLT phase shifters operate on picosecond timescales and support electro-optic bandwidths exceeding gigahertz, indicating that the demonstrated tracking speed is not a fundamental material constraint but a control-architecture limit that can be further relaxed with faster electronics.

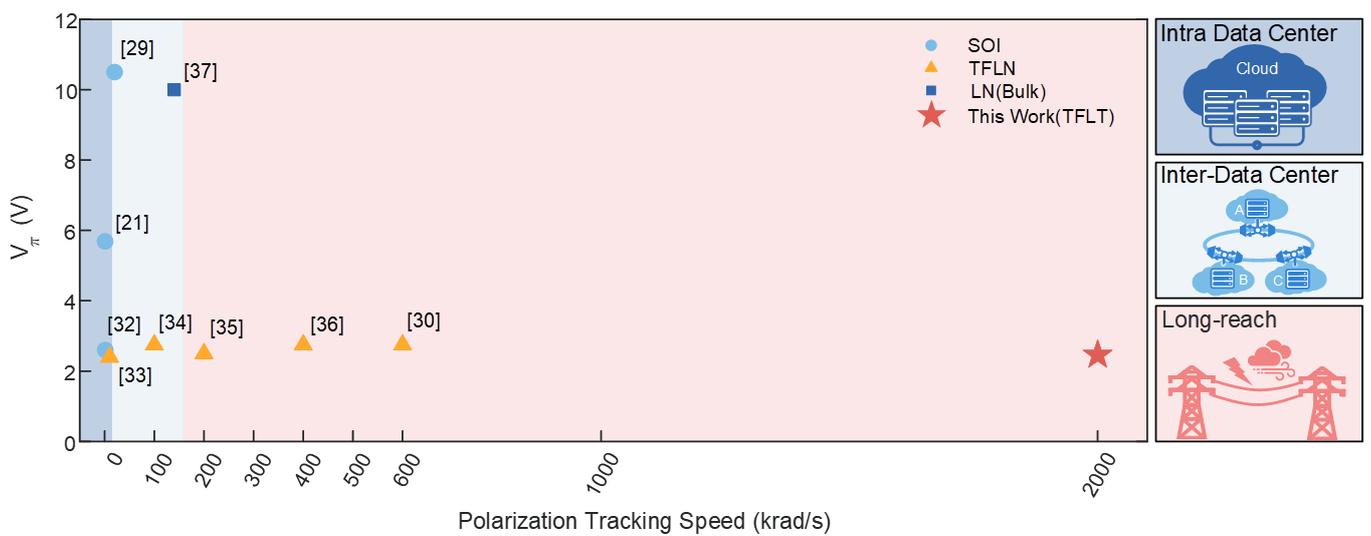

**Figure 7. Performance comparison of representative APCS.** Comparison of polarization tracking speed and half-wave

voltage across representative integrated platforms, with shaded regions indicating application-relevant tracking-speed regimes.

Figure 7 captures this advantage from a device-level perspective by comparing polarization tracking speed and half-wave voltage across representative integrated platforms reported in the literature[21,29-30,32-37]. Specifically, the dark-blue region corresponds to tracking speeds about 1 krad·s$^{-1}$, which are generally sufficient for intra–data-center links with relatively slow polarization variations. Platforms in this regime are predominantly based on silicon photonic platforms employing thermo-optic phase shifters[21,32], where the intrinsic thermal response fundamentally limits the achievable tracking speed. The light-blue region represents tracking speeds below 100 krad·s$^{-1}$, relevant to inter–data-center transmission where polarization dynamics are more pronounced. This regime is primarily populated by TFLN platforms leveraging electro-optic modulation[33-34]. In contrast, the magenta region denotes tracking speeds exceeding 100 krad·s$^{-1}$, enabling robust operation in more demanding and broadly applicable scenarios characterized by rapid and continuous SOP evolution. State-of-the-art TFLN-based implementations reach tracking speeds of up to 600 krad·s$^{-1}$[30], whereas our work based on TFLT extends this capability to the Mrad·s$^{-1}$ regime, achieving tracking speeds up to 2 Mrad·s$^{-1}$. In this two-dimensional space, the present TFLT-based APC occupies a unique Pareto-optimal position, simultaneously achieving Mrad·s$^{-1}$ tracking and sub-2.5-V operation. This combination is difficult to achieve on other widely used integrated platforms, where either thermal phase shifters, higher drive voltages, or long-term bias management typically impose trade-offs between speed, efficiency, and stability. Nevertheless, high tracking speed and low drive voltage alone do not guarantee robust operation under continuously varying polarization, where phase-boundary effects become the dominant limitation.

However, despite these device-level advances, most existing APC implementations remain unable to sustain reset-free operation under continuously varying SOPs. Some approaches employ optical state switching to prevent the control phase from reaching boundary regions[22], thereby avoiding phase resets. While effective in principle, this strategy introduces additional optical switching elements, increases power consumption, and still suffers from transient discontinuities during state transitions. Alternatively, other schemes modify the tracking target, for example by constraining the SOP to the $S_1 = 0$ circle on the Poincaré sphere[20,30]. Although this avoids the full 0–2π phase reset, it introduces discontinuous ±π phase jumps and incurs an intrinsic 3 dB power penalty at the output, limiting its suitability for power-sensitive communication systems.

These limitations underscore the system-level importance of achieving truly reset-free polarization control. In both SHC and DP-IMDD links, polarization-induced fluctuations directly translate into power imbalance, crosstalk, and burst errors that cannot be corrected in real time by digital signal processing. Our measurements show that when polarization dynamics exceed the control bandwidth, demodulation fails not because polarization tracking collapses, but because the finite phase space of the phase shifter becomes exhausted. In all realistic fiber environments, however, polarization dynamics remain well below this extreme boundary, allowing the proposed architecture to maintain stable, endless operation without resets.

From a platform perspective, TFLT occupies a unique position among integrated photonic technologies. It combines ultrafast electro-optic response with intrinsically low DC drift, eliminating the need for auxiliary thermal bias control that complicates and destabilizes other platforms. This enables a class of feedback-driven photonic systems in which speed, stability, and scalability coexist—an essential requirement for next-generation optical interconnects in AI-driven data centers.

Together, these results demonstrate that ultrafast, reset-free polarization control is not limited by photonic materials but by control-loop architecture, and that by co-designing device physics with boundary-aware algorithms, integrated photonics can enter a new dynamic regime previously inaccessible to practical communication systems.

## Conclusion

We have demonstrated the first integrated polarization controller based on thin-film lithium tantalate, achieving reset-free phase-jump-free polarization tracking at unprecedented Mrad·s$^{-1}$ speeds. By combining the intrinsically fast and drift-free electro-optic response of the TFLT platform with a finite-boundary gradient-descent control algorithm, we realize continuous polarization stabilization without phase resets and jumps, even under extreme and rapidly varying SOP.

The proposed device exhibits low PDL, low half-wave voltage, and long-term bias stability, while supporting polarization dynamics that exceed those encountered in realistic fiber environments by orders of magnitude. When integrated into a 400-Gbps SHC transmission system, the APC maintains error-free operation under lightning-scale polarization scrambling, directly translating ultrafast optical control into system-level robustness. The observed performance boundary is shown to arise from finite control-loop latency rather than intrinsic photonic limitations, indicating clear pathways for further scaling through electronic and algorithmic optimization.

Beyond SHC links, the demonstrated platform is equally applicable to emerging DP-IMDD architectures, where stable and high-speed polarization control is essential for preserving channel orthogonality and maximizing spectral efficiency. By unlocking a previously inaccessible regime of ultrafast, low-power, and reset-free polarization control on a scalable integrated platform, this work establishes TFLT as a key enabler for next-generation optical interconnects in AI-driven data-center and short-reach communication networks.

## Methods

**Ultrafast polarization control electronics**

To meet the demands of ultrafast polarization control, the system architecture was designed to support both high-bandwidth phase shifter and stable closed-loop operation. Real-time polarization tracking was implemented on a FPGA, providing the deterministic latency and computational throughput required for high-speed feedback control (Figure 8).

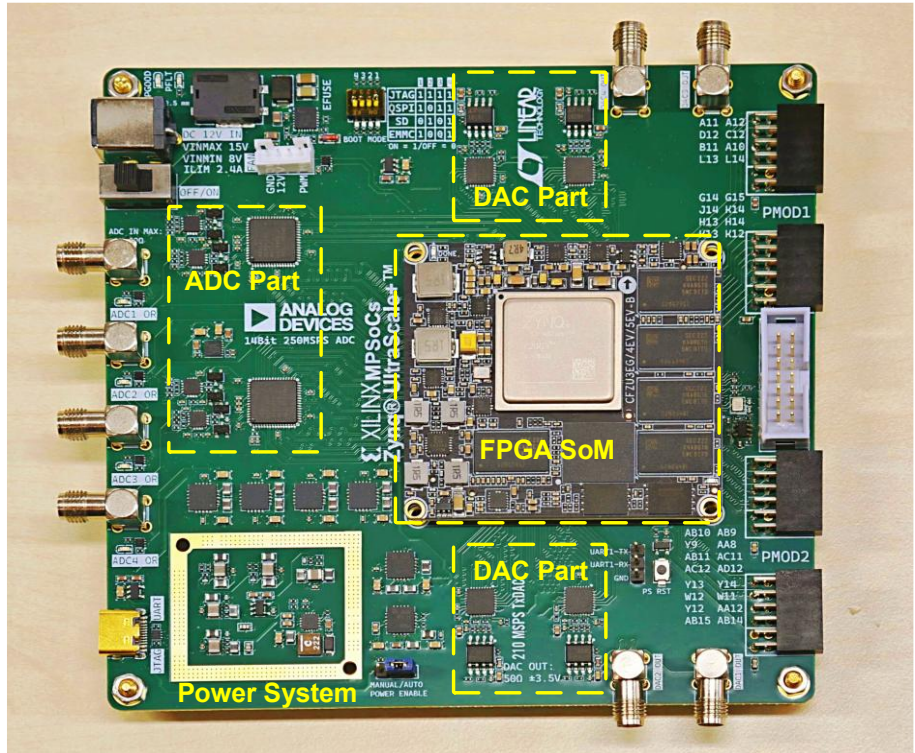

**Figure 8. Architecture of the polarization control electronics.**

Real-time operation relies on a high-speed data-conversion subsystem for acquiring polarization-dependent error signals and generating the corresponding control voltages. As shown in Figure 8, the subsystem integrates multi-channel analog-to-digital converters (ADCs) and digital-to-analog converters (DACs) with dedicated analog front-end and back-end circuitry. The achievable polarization-control performance is primarily constrained by the quantization resolution and conversion latency of the data converters. Sufficient resolution is required to preserve sensitivity to small SOP deviations and to enable fine-grained polarization adjustment without quantization noise limiting the effective extinction ratio of the polarization-control elements. Conversion latency, in contrast, directly sets the upper bound of the closed-loop bandwidth. For the ADCs, the dominant latency originates from pipeline quantization and the associated digital interface, which increases with sampling rate. For the DACs, latency is mainly determined by the digital interface and output settling dynamics. Balancing these constraints, 14-bit ADCs and 14-bit DACs operating at 200 MSPS were employed, resulting in a total loop latency of approximately 100 ns.

Beyond data-converter resolution and loop latency, the stability of ultrafast polarization control is critically conditioned by the power integrity of the control electronics. Under well-shielded experimental conditions, the dominant noise sources arise from intrinsic device noise and power-supply–related interference rather than external electromagnetic coupling. Owing to the finite power-supply rejection ratio (PSRR) of both analog and digital integrated circuits, noise on the supply rails can couple into sensitive signal paths, degrading signal fidelity and compromising closed-loop stability. As illustrated in Figure 8, the control system incorporates multiple supply domains to support digital processing, data conversion, and analog front-end circuitry. High-efficiency voltage conversion, including voltage inversion and step-up/step-down regulation implemented using switching regulators, introduces broadband conducted noise and high-frequency switching transients. These disturbances can further propagate through shared impedance in the power-distribution network or via parasitic coupling between supply domains. To suppress these effects, the power-management system was implemented using a hierarchical DC–DC plus low-dropout regulator (LDO) architecture, with electrically and physically isolated supply domains for analog and digital circuits. The DC–DC stages provide efficient coarse regulation, while the LDO stages locally attenuate residual switching ripple and high-frequency noise. In addition, on-board electromagnetic shielding was applied to the DC–DC regulator circuitry to suppress radiated switching noise and prevent parasitic coupling into adjacent analog signal paths. This architecture preserves high power-conversion efficiency while providing the low-noise supply rails required for stable high-speed closed-loop polarization control.

**FPGA implementation of the FBGD algorithm**

In the polarization-tracking algorithm, the input is the feedback optical power digitized by the ADC, and the output is the updated control voltage generated by the DAC to drive the phase shifters. The execution flow of the algorithm, illustrated in Figure 4c, is implemented on the FPGA using a finite-state machine to ensure deterministic timing and real-time operation.

Each control cycle is partitioned into four sequential states. In the first state, controlled perturbations are applied to the DAC outputs to probe the local response of the loss function. The second state compensates for the loop latency, allowing the combined optical and electronic system to reach a steady response. In the third state, the feedback signal is sampled to estimate the partial derivative of the loss function with respect to the control phase, as defined in Eq. (3). In the final state, the control voltage is updated according to the computed gradient. Continuous polarization tracking is achieved by repeatedly executing this four-state cycle, enabling deterministic, low-latency, real-time implementation of the FBGD algorithm on the FPGA.

# Reference

1. Singh, A. *et al.* Jupiter Rising: A Decade of Clos Topologies and Centralized Control in Google's Datacenter Network. *ACM SIGCOMM Comput. Commun. Rev.* **45**, 183–197 (2015).

2. Kumar, A. V. *et al.* A case for server-scale photonic connectivity. in *Proceedings of the 23rd ACM Workshop on Hot Topics in Networks* 290–299 (ACM, Irvine CA USA, 2024).

3. Sun, A. *et al.* Edge-guided inverse design of digital metamaterial-based mode multiplexers for high-capacity multi-dimensional optical interconnect. *Nat. Commun.* **16**, 2372 (2025).

4. Wade, M. *et al.* An Error-free 1 Tbps WDM Optical I/O Chiplet and Multi-wavelength Multi-port Laser. in *Optical Fiber Communication Conference (OFC) 2021 (2021), paper F3C.6* (Optica Publishing Group, 2021).

5. Cheng, J. *et al.* A Low-Complexity Adaptive Equalizer for Digital Coherent Short-Reach Optical Transmission Systems. in *Optical Fiber Communication Conference (OFC) 2019* M3H.2 (OSA, San Diego, California, 2019).

6. Tang, M. *et al.* Self-Homodyne Coherent Systems for Short-Reach Optical Interconnects. in *2023 Optical Fiber Communications Conference and Exhibition (OFC)* 1–3 (2023).

7. Li, W. *et al.* 100-km Polarization-Orthogonal Self-Homodyne Coherent WDM Transmission Using

Nonlinearity Suppressed SOA. in *CLEO 2023* STu4G.5 (Optica Publishing Group, San Jose, CA, 2023).

8. Zhang, M. *et al.* First Baud-Rate Sampled DSP-Free Self-Homodyne Coherent Receiver. in *2023 Asia Communications and Photonics Conference/2023 International Photonics and Optoelectronics Meetings (ACP/POEM)* 1–4 (IEEE, Wuhan, China, 2023).

9. Ji, H. *et al.* Complementary Polarization-Diversity Coherent Receiver for Self-Coherent Homodyne Detection With Rapid Polarization Tracking. *J. Light. Technol.* **40**, 2773–2779 (2022).

10. Walker, N. G. *et al.* Polarization control for coherent communications. *J. Light. Technol.* **8**, 438–458 (1990).

11. Pittalà, F. *et al.* Laboratory Measurements of SOP Transients due to Lightning Strikes on OPGW Cables. in *2018 Optical Fiber Communications Conference and Exposition (OFC)* 1–3 (2018).

12. Charlton, D. *et al.* Field measurements of SOP transients in OPGW, with time and location correlation to lightning strikes. *Opt. Express* **25**, 9689 (2017).

13. Cruzes, S. Failure Management Overview in Optical Networks. *IEEE Access* **12**, 169170–169193 (2024).

14. Zhao, Y. *et al.* Dual-Polarization IMDD System for Data-Center Connectivity. *J. Light. Technol.* **43**, 6044–6049 (2025).

15. Zhao, Y. et al. 425-Gbps/λ Dual-polarization IMDD Transceiver. Optica (OFC) 2025.

16. Rahim, A. *et al.* Expanding the Silicon Photonics Portfolio With Silicon Nitride Photonic Integrated Circuits. *J. Light. Technol.* **PP**, 1–1 (2016).

17. Tong, W. *et al.* An Efficient, Fast-Responding, Low-Loss Thermo-Optic Phase Shifter Based on a Hydrogen-Doped Indium Oxide Microheater. *Laser Photonics Rev.* **17**, 2201032 (2023).

18. Shi, J. *et al.* Alleviation of DC drift in a thin-film lithium niobate modulator utilizing $Ar^+$ ion milling. *Opt. Lett.* **50**, 1703 (2025).

19. Ying, P. *et al.* Low-loss edge-coupling thin-film lithium niobate modulator with an efficient phase shifter. *Opt. Lett.* **46**, 6 (2021).

20. Oswald, P. *et al.* Deterministic analysis of endless tuning of polarization controllers. *J. Light. Technol.* **24**, 2932–2939 (2006).

21. Wang, X. *et al.* Reset-free adaptive polarization controller on a silicon-photonic platform for a self-coherent communication system. *Opt. Lett.* **48**, 1546–1549 (2023).

22. Doerr, C. R. *et al.* PDM-DQPSK Silicon Receiver With Integrated Monitor and Minimum Number of Controls. *IEEE Photonics Technol. Lett.* **24**, 697–699 (2012).

23. Hu, C. *et al.* High-efficient coupler for thin-film lithium niobate waveguide devices. *Opt. Express* **29**, 5397 (2021).

24. Holzgrafe, J. et al. Relaxation of the electro-optic response in thin-film lithium niobate modulators. *Opt. Express* **29**, 3619–3631 (2024).

25. Yeh, M. *et al*. Interface-mediated dc electro-optic instability in lithium niobate nanophotonics. *Optica*(2025).

26. Grachev, V. G. & Malovichko, G. I. Structures of impurity defects in lithium niobate and tantalate derived from electron paramagnetic and electron nuclear double resonance data. *Crystals* **11**, 339 (2021).

27. Wang, H. *et al*. Optical switch with an ultralical low DC drift based on thin-film lithium tantalate. *Opt. Lett.* **49**, 5019–5022 (2024).

28. Shi, J. *et al.* Alleviation of DC drift in a thin-film lithium niobate modulator utilizing Ar$^+$ ion milling. *Opt. Lett.* **50**, 1703 (2025).

29. Wang, W. *et al.* CMOS-compatible high-speed endless automatic polarization controller. *APL Photonics* **9**, 066116 (2024).

30. Chen, W. *et al.* 600-krad/s Polarization Tracking Using a DC-Stable Integrated Polarization Controller Based on Thin Film Lithium Niobate. in *ECOC 2024; 50th European Conference on Optical Communication* 735–737 (2024).

31. Nespola, A. *et al.* Proof of Concept of Polarization-Multiplexed PAM Using a Compact Si-Ph Device. *IEEE Photonics Technol. Lett.* **31**, 62–65 (2019).

32. Chang, P.-H. *et al.* A Monolithic Polarization Controller and Tracking Loop for Optical Interconnect Demonstrated on a 90 nm Silicon CMOS-Photonic Platform. *J. Light. Technol.* **41**, 3832–3841 (2023).

33. Lin, Z. *et al.* High-performance polarization management devices based on thin-film lithium niobate. *Light Sci. Appl.* **11**, 93 (2022).

34. Li, Z. *et al.* High-speed polarization tracking using thin film lithium niobate integrated dynamic polarization controller. *Opt. Express* **31**, 39369 (2023).

35. Ye, J. *et al.* 200-krad/s Polarization Tracking Using Thin Film Lithium Niobate Integrated Dynamic Polarization Controller. in *2024 IEEE Opto-Electronics and Communications Conference (OECC)* 1–3 (2024).

36. Guo, C. *et al.* Integrated Self-coherent Optical Transceivers for Data Center Applications Based on Thin-Film Lithium Niobate Platform. *J. Light. Technol.* 1–7 (2025).

37. Koch, B. *et al.* 140-krad/s, 254-Gigaradian Endless Optical Polarization Tracking, Independent of

Analyzed Output Polarization. in *Optical Fiber Communication Conference* OTu1G.6 (OSA, Los Angeles, California, 2012).

**Acknowledgements**

This work is supported by the National Key Research and Development Program of China (No. 2025YFA1212900), the National Natural Science Foundation of China (No. 62205114，No. 62225110，No. 62175079 and No. 62205119), Quantum Science and Technology-National Science and Technology Major Project 2021ZD0300701, Key R&D Program of Shandong Province, China (2023CXGC010110), Hubei Optical Fundamental Research Centre and Advanced Micro-Nano Fabrication Center, HUST. Lithium tantalate device fabrication is performed at the cleanroom of Wuhan ANPI Corporation. We thank Xianda Zheng from Changchun Institute of Optics, Fine Mechanics and Physics, CAS for technical support.

**Author contributions**

S.Y., C.Z., J.X., M.T., conceived and planned the research. S.L. and G.Y. designed and fabricated the devices; Z.G. designed and implemented the control system and realized the control algorithm proposed by M.Z.; Z.G. and G.Y. led the measurement and analysis with assistance from C.Z. and M.Z.; Z.G., S.L. and G.Y. wrote the paper with inputs from all the co-authors. The work was supervised by S.Y., C.Z., J.X., M.T.

**Competing interests**

The authors declare no competing interests.

**Additional information**

**Supplementary information** The online version contains supplementary material available at https://doi.org/xxxxxx.